\documentclass[
aps,
prl,
twocolumn,
superscriptaddress,
groupaddress,
]{revtex4}

\usepackage{graphicx}
\usepackage{dcolumn}
\usepackage{amsmath}

\def\eac{\epsilon_{\mbox{{\scriptsize ac}}}}
\def\edc{\epsilon_{\mbox{\scriptsize dc}}}

\def\oc{\omega_{\mbox{\scriptsize {C}}}}
\def\oh{\omega_{\mbox{\scriptsize {H}}}}

\def\rc{R_{\mbox{\scriptsize {C}}}}

\def\tq{\tau_{\mbox{\scriptsize {q}}}}

\begin{document}
\title{
Multiphoton processes at cyclotron resonance subharmonics in a 2D electron system under DC and microwave excitation
}
\author{S. Chakraborty}
\affiliation{National High Magnetic Field Laboratory, Tallahassee, Florida 32310, USA} 
\author{A.\,T. Hatke}
\affiliation{National High Magnetic Field Laboratory, Tallahassee, Florida 32310, USA} 
\author{L.\,W. Engel}
\affiliation{National High Magnetic Field Laboratory, Tallahassee, Florida 32310, USA} 

\author{J. D. Watson}
\affiliation{Department of Physics, Purdue University, West Lafayette, Indiana 47907, USA} 
\affiliation{Birck Nanotechnology Center, School of Materials Engineering and School of Electrical and Computer Engineering, Purdue University, West Lafayette, Indiana 47907, USA}
\author{M. J. Manfra}
\affiliation{Department of Physics, Purdue University, West Lafayette, Indiana 47907, USA} 
\affiliation{Birck Nanotechnology Center, School of Materials Engineering and School of Electrical and Computer Engineering, Purdue University, West Lafayette, Indiana 47907, USA}

\received{\today}

\begin{abstract}
We investigate a two-dimensional electron system (2DES) under microwave illumination at cyclotron resonance subharmonics.
The 2DES carries sufficient direct current, $I$, that the differential resistivity oscillates as $I$ is swept.
At magnetic fields sufficient to resolve individual Landau levels, we find the number of oscillations within an $I$ range systematically changes with increasing microwave power.
Microwave absorption and emission of $N$ photons, where $N$ is controlled by the microwave power, describes our observations in the framework of the displacement mechanism of impurity scattering between Hall-field tilted Landau levels.
\end{abstract} 

\maketitle

Low magnetic field $(B)$ transport of two dimensional electron systems (2DESs) under nonequilibrium conditions has been a subject of intense study for over a decade.
Illumination with microwave (ac) radiation results in $1/B$-periodic resistance oscillations termed microwave induced resistance oscillations (MIROs) \citep{zudov:2001a,ye:2001,mani:2002,dmitriev:2012}.
Additionally, application of sufficiently large direct current (dc) in a Hall bar geometry results in $1/B$-periodic differential resistivity $(r)$ oscillations called Hall-field induced resistance oscillations (HIROs) \citep{yang:2002,zhang:2007a,hatke:2009c}.
MIROs are periodic in $\eac=\omega/\oc$, where $\omega=2\pi f$ is the frequency of the microwave radiation and $\oc$ is the cyclotron frequency.
MIRO maximum-minimum pairs occur symmetrically offset from integer values of $\eac$ at $\eac^{\pm}=m\mp \phi$, where $m$ is an integer and $\phi$ is referred to as the phase \citep{mani:2004d,studenikin:2005,hatke:2011f,hatke:2011e}.
HIROs are periodic in $\edc=\oh/\oc$, where $\oh=\sqrt{8\pi/n_{e}}I/we$ with $n_{e}$ the carrier density, $I$ the applied direct current, and $w$ the Hall bar width.
Maxima and minima of HIROs occur, respectively, at $\edc=m$ and $\edc=m-1/2$ \citep{yang:2002}. 
Both oscillation types have been investigated theoretically considering two main mechanisms:
the first, called the displacement mechanism \citep{ryzhii:1970,durst:2003,vavilov:2004}, is based on the modification of impurity scattering in the presence of ac or dc excitation, and the second, inelastic mechanism \citep{dmitriev:2005,dmitriev:2007,dmitriev:2009b}, is based on a nonequilibrium density of states. 
The inelastic and displacement mechanisms are argued to be valid for different experimental conditions depending on whether 1) the magnetic field is sufficiently large for the Landau levels (LLs) to be well separated, $\oc\tq>\pi/2$ where $\tq$ is the quantum scattering time, or 2) on whether there is dc excitation.

For large microwave intensity, MIROs can occur at rational fractions of $\eac$, with $\eac=1/m$ the most readily obtainable experimental series \citep{dorozhkin:2003,willett:2004,zudov:2004,zudov:2006a,dorozhkin:2007,wiedmann:2009a}.
These fractional MIROs are a result of multiphoton processes.
In the displacement mechanism, transitions can occur due to sequential absorption of single photons through real intermediate states \citep{lei:2006a,pechenezhskii:2007}, while in the inelastic mechanism, transitions occur through intermediate virtual states \citep{dmitriev:2007b}.
An inelastic model theory \citep{dmitriev:2007b} of fractional MIROs in the separated-LL regime incorporates virtual transitions between microwave-induced sidebands in the density of states and predicts the inelastic mechanism to overwhelm the displacement mechanism.

In this Letter we study samples with applied direct current \citep{vavilov:2007}, which is expected to suppress the inelastic mechanism relative to the displacement mechanism for $2\pi\edc\gg1$ \citep{khodas:2008,lei:2009}.
Working in the regime of separated LLs, we perform a combined (ac + dc) \citep{zhang:2007c,zhang:2008b,hatke:2008a} experiment, in which we subject the sample to direct current and to microwaves at fractional-$\eac$ using a coplanar waveguide structure. 
We find that the number of oscillations within an $\edc$ range changes systematically with increasing microwave power in a way that clearly depends on the number of participating photons.
Our results can be described in terms of competition between scattering events involving different numbers of photons.

\begin{figure}[t]
\vspace{-0.1 in}
\includegraphics[width=0.475\textwidth]{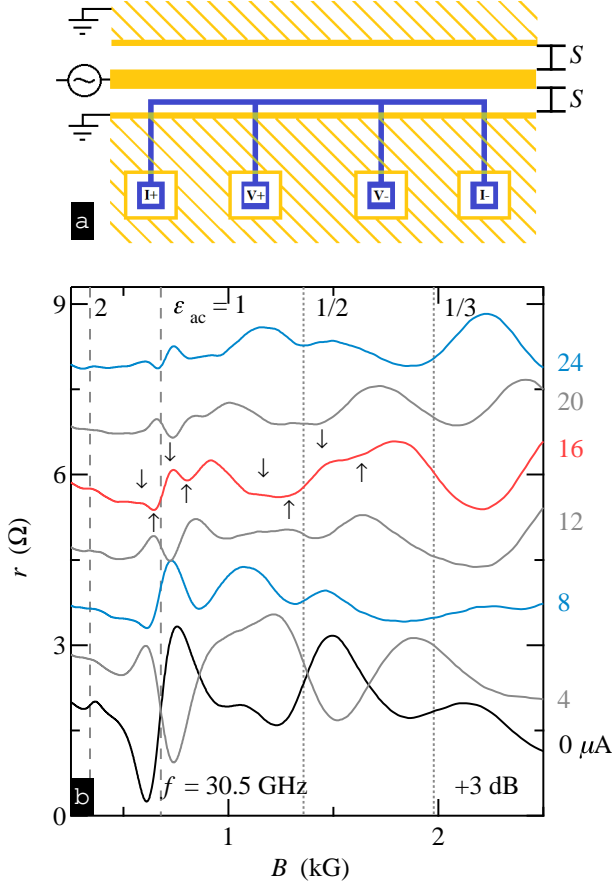}
\vspace{-0.2 in}
\caption{(color online) (a) Schematic of the microwave set up for a top view, not to scale.
(b) $r$ vs $B$ with microwave radiation of $f=30.5$\,GHz at different applied currents from $I=0$ to $24\,\mu$A in steps of $4\,\mu$A for microwave power $+3\,$dB.
Vertical dashed lines denote the cyclotron resonance and its second harmonic and the dotted lines mark its second and third subharmonic.
Traces are vertically offset for clarity.
}
\label{fig1}
\vspace{-0.2 in}
\end{figure}

The microwave setup is shown schematically in Fig.\,\ref{fig1}\,(a).
We lithographically defined a Hall bar of width $w=20\,\mu$m, etched from a symmetrically doped GaAs/AlGaAs quantum well, and deposited Ge/Au/Ni contacts.
A coplanar waveguide \citep{wen:1969} with slot width $S=100\,\mu$m, defined as the distance between the driven center conductor and the ground plane, was superimposed on the sample surface with the Hall bar oriented along the slot.
The slot confined the microwave electric field.
The contacts were located $\sim 6 S$ from the slot and were shielded by the ground plane.
The density, $n\simeq 3.7 \times 10^{11}$ cm$^{-2}$, and mobility, $\mu\simeq 5.7 \times 10^{6}$ cm$^{2}$/Vs, were obtained by brief illumination with a red light emitting diode at a few Kelvin.
The sample was mounted on a brass block and was kept at $T=1.4\,$K in vacuum for the measurements.
The differential resistivity, $r\equiv dV/dI$, was measured with a lock-in amplifier at a few Hz. 
In this paper, $0\,$dB corresponds to a microwave rms voltage of $\sim 4.3\,$mV on the center conductor.

In Fig.\,\ref{fig1}\,(b) we plot $r$ vs $B$ obtained while driving the transmission line at frequency $f=30.5\,$GHz, power $+3\,$dB, and fixed currents from $I=0$ to $24\,\mu$A in steps of $4\,\mu$A. 
For the $I=0\,\mu$A trace $\phi<1/4$ at $\eac=2$, which is consistent with a high power regime of previous work \citep{note_highp}. 
Additionally, fractional MIROs centered at $\eac=1/2$ and $1/3$ are observed. 
For $I=4\,\mu$A, at $\eac=1$ ($\edc\sim 1/2$) and $\eac=1/2$ ($\edc\sim 1/4$), the oscillations have flipped.
Maxima at $4\,\mu$A replace minima at $0\,\mu$A while minima at $4\,\mu$A replace maxima at $0\,\mu$A. 
At $I=8\,\mu$A the $\eac=1$ ($\edc\sim1$) and $\eac=1/2$ ($\edc\sim 1/2$) oscillation extrema again match those in the $I=0\,\mu$A trace.  
For $I>8\,\mu$A additional maxima and minima are observed near $\eac=1$ and $\eac=1/2$. 
On the $I=16\,\mu$A trace these extrema are marked by $\downarrow$ and $\uparrow$. 
The additional maximum and minimum near $\eac=1$ agree with previous high microwave power and strong direct current observations \citep{khodas:2010}.
In what follows we restrict our measurements to the separated LL regime.

\begin{figure}[t]
\vspace{-0.1 in}
\includegraphics[width=0.48\textwidth]{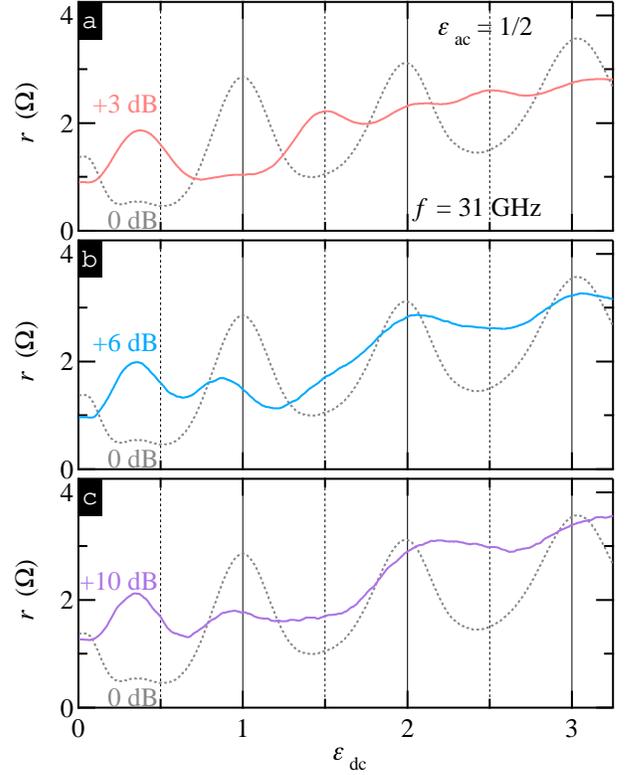}
\vspace{-0.2 in}
\caption{(color online) 
$r$ vs $\edc$ at $f=31\,$GHz and $\eac=1/2$ for a baseline weak power, dotted trace in each panel, and increasing power of $+3\,$dB (a), $+6\,$dB (b), and $+10\,$dB (c).
}
\label{half}
\vspace{-0.2 in}
\end{figure}

\begin{figure}[t]
\vspace{-0.1 in}
\includegraphics[width=0.48\textwidth]{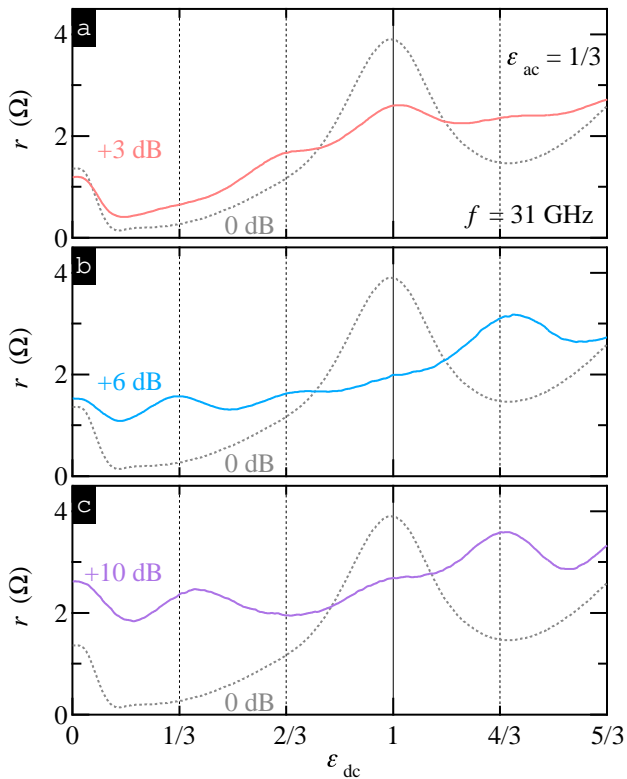}
\vspace{-0.2 in}
\caption{(color online) 
$r$ vs $\edc$ at $f=31\,$GHz and $\eac=1/3$ for a baseline weak power, dotted trace in each panel, and increasing power of $+3\,$dB (a), $+6\,$dB (b), and $+10\,$dB (c).}
\label{third}
\vspace{-0.2 in}
\end{figure}

In Figs.\,\ref{half} and \ref{third} we show $r$ vs $\edc$ taken by sweeping $I$ at fixed $B$.
Figure\,\ref{half} shows data for $\eac=1/2$ and $f=31\,$GHz. 
A $0\,$dB trace (dotted line) is presented as a baseline in each plot.
The $0\,$dB trace contains several well-defined HIROs up to $\edc=3$, with maxima at $\edc=m$ and minima at $\edc=m-1/2$. 
In Fig.\,\ref{half}\,(a) we show the effect of increasing the microwave power by $+3\,$dB. 
Here, maxima are observed near $\edc=1/2,3/2$ and $5/2$ \citep{ac_dc_note}.
In addition, maxima at $\edc=m$ are still present, though with reduced amplitude compared to the $0\,$dB trace, hence the entire $+3\,$dB $r$ vs $\edc$ trace in Fig.\,\ref{half}\,(a) shows a microwave power induced doubling of the $\edc$-frequency for up to \textit{six} oscillations \citep{doubling_note}.

Figure\,\ref{half}\,(b) shows data taken with an increased microwave power of $+6\,$dB. 
The strength of the $\edc=1/2$ maximum is little changed, though the $\edc=3/2$ maximum is almost completely suppressed. 
The maximum at $\edc=5/2$ has become a minimum, now more closely resembling the $0\,$dB trace than the $+3\,$dB trace. 
In addition, the amplitudes of the maxima at $\edc=m$ are increased relative to their values at $+3\,$dB. 
For the highest microwave power, $+10\,$dB, as shown in Fig.\,\ref{half}\,(c), $\edc=3/2$ is a minimum. 
Throughout the power dependence shown in Fig.\,\ref{half}, the $\edc=1/2$ maximum persists with only weak amplitude variation. 

The same measurements except at $\eac=1/3$ are presented in Fig.\,\ref{third}. 
The baseline, $0\,$dB, trace shows only a single HIRO maximum at $\edc=1$ and a strong zero bias peak \citep{zhang:2007b,hatke:2012d}.
In Fig.\,\ref{third}\,(a) the $+3\,$dB trace shows a weakening of the $\edc=1$ maximum relative to the one at $0\,$dB and two additional maxima at $\edc=2/3$ and $4/3$.
For the $+6\,$dB trace, oscillation maxima are present at $\edc=1/3,2/3,1\,$and $4/3$, with the $\edc=1$ maximum quite weak.
The presence of these maxima is consistent with \textit{tripling} of the $\edc$-frequency.
At the largest power of $+10\,$dB, in Fig.\,\ref{third}\,(c), the $\edc=1/3$ and $\edc=4/3$ maxima become stronger and the $\edc=2/3$ maximum of lower power has changed to a minimum.

The main result of this paper is that the observed change of the $\edc$-frequency with microwave power is compactly described within the displacement mechanism picture by scattering processes that involve absorption and emission of varying photon numbers.
In the separated LL regime the condition for maxima in $r$ due to a combined (ac + dc) transition involving arbitrary photon number is 
\begin{equation}
\edc+N\eac=m
\label{eq}
\end{equation}
with $N=0,\pm1,\pm2...$ where the positive (negative) sign denotes photon absorption (emission) and $m$ is a nonnegative integer.
The equation was originally used \citep{lei:2009} to describe the $\edc$-frequency doubling observations \citep{hatke:2008b}.

We now show how Eq.\,\ref{eq} describes our data as increase of the microwave power changes $N$ and hence the $\edc$-frequency.
At $\eac=1/2$ combined (ac + dc) transitions involving even $N$ produce maxima for $\edc=m$, while transitions with odd $N$ produce maxima at $\edc=m-1/2$.
The $0\,$dB trace in Fig.\,\ref{half} reflects weak $N=1$ processes that compete with $N=0$ processes. 
This interpretation is confirmed by the presence of a slight maximum at $\edc=1/2$, an $N=1$ transition, while $N=0$ processes produce minima at $\edc=3/2$ and $5/2$. 
With increasing power, $N=1$ transitions produce maxima at $\edc=3/2$ and $5/2$ and also produce minima at $\edc=m$, whose effect is to decrease the amplitude of maxima generated by $N=0$ processes.
However, the continued presence of $\edc=m$ maxima, for example in the $+3\,$dB trace of Fig.\,\ref{half}\,(a), means the number of participating photons is not restricted to a single value of $N$. 
The observed loss of the $\edc=5/2$ maximum and then of the $\edc=3/2$ maximum with increasing power is due to $N=2$ processes, which strengthen $\edc=m$ maxima and weaken $\edc=m-1/2$ maxima.
The $\edc=1/2$ maximum for $+10\,$dB shows that $N=1$ processes remain in competition with $N=2$ processes.

\begin{figure}[t]
\vspace{-0.1 in}
\includegraphics[width=0.48\textwidth]{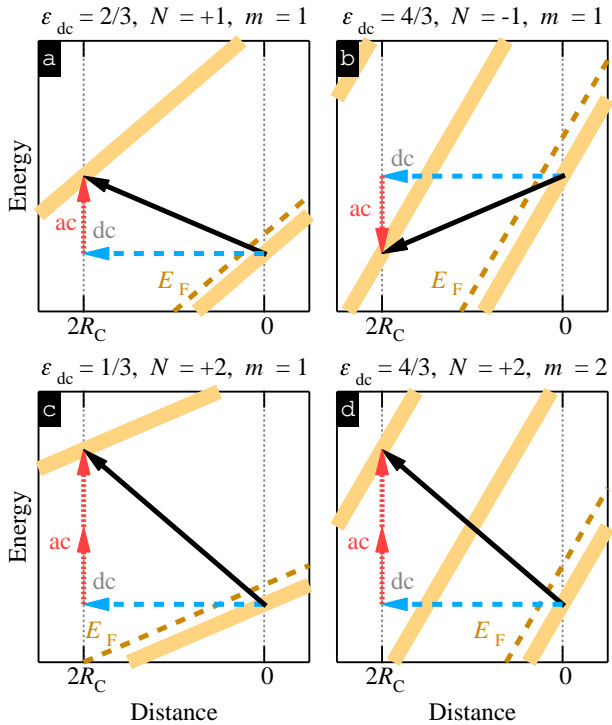}
\vspace{-0.2 in}
\caption{(color online) 
Energy vs distance Landau level diagrams for $\eac=1/3$ for maxima at (a) $\edc=2/3$, $N=1$; (b) $\edc=4/3$, $N=-1$; (c) $\edc=1/3$, $N=2$; and (d) $\edc=4/3$, $N=2$.
Thick lines denote Hall-field tilted Landau levels, vertical dotted arrows represent energy change due to absorption (up arrows) or emission (down arrows) of a photon, horizontal dashed arrows for spatial change of an electron backscattering by the cyclotron diameter $(2\rc)$, and inclined solid arrows for a combined transition.
}
\label{lls}
\vspace{-0.2 in}
\end{figure}

In Fig.\,\ref{lls} we use energy-space LL diagrams \citep{hatke:2008a,hatke:2008b} to illustrate the conditions for maxima in terms of Eq.\,\ref{eq}.
The $\eac=1/3$ results of Fig.\,\ref{third} are shown for $N=1$ at $\edc=2/3$ (a) and $\edc=4/3$ (b) and $N=2$ at $\edc=1/3$ (c) and $\edc=4/3$ (d). 
Thick lines denote well separated-LLs that are tilted by the Hall field, horizontal dashed arrows represent transitions from impurity scattering, vertical dotted arrows are microwave transitions (up for absorption and down for emission), and inclined solid arrows are combined (ac + dc) transitions.
Maxima occur when a transition terminates at the center of a LL.
The $+3\,$dB trace in Fig.\,\ref{third}\,(a) shows $N=\pm1$ processes are operational by maxima at $\edc=2/3$ and $\edc=4/3$, while the $\edc=1$ maximum demonstrates that $N=0$ processes remain.
For the $+6\,$dB trace in Fig.\,\ref{third}\,(b) a combination of $N=\pm1$ and $N=2$ processes are present; $N=1$ processes produce a maximum at $\edc=2/3$, $N=2$ processes produce a maximum at $\edc=1/3$, and both $N=-1$ and $N=2$ contribute to the maximum at $\edc=4/3$.
For $\edc=4/3$ the $N=2$ processes occur for absorption and scattering but the $N=-1$ process results in a maximum only if photon \textit{emission} occurs, and $N=+1$ processes do not terminate at the center of a LL.
For the $+10\,$dB trace in Fig.\,\ref{third}\,(c) $N=2$ processes become more important than $N=1$ processes at $\edc=2/3$ so that a minimum is formed, instead of the maximum seen at $+3\,$dB and $+6\,$dB. 

Though Eq.\,\ref{eq} compactly describes our observations, the numerical results of Ref.\,\citep{lei:2009} do not discuss the role of microwave power.
For the separated LL regime, Ref.\,\citep{lei:2009} reproduced the experimentally observed $\edc$-frequency doubling of Ref.\,\citep{hatke:2008b} and predicted a rough $\edc$-frequency tripling at $\eac=1/3$.
An extension of \citep{lei:2009} to an analytic theory \citep{lei:2011}, using the overlapping LL simplification, failed to produce the additional oscillations near $\eac=1/2$ in Fig.\,\ref{fig1}\,(b) and did not show the presently reported $\edc$-frequency change at fractional MIROs.

Previous works do not adequately predict our observations, which are explainable within the displacement mechanism framework of Eq.\,\ref{eq}.
Reference\,\citep{zudov:2009} advanced a displacement description of the inversion of extrema observed in $r$ vs $I$ at $\eac=3/2$ for overlapping LLs at large microwave power, but that picture does not match our observed $\edc$-frequency change at $\eac=1/2$ or $\eac=1/3$.
Lastly, a theory \citep{khodas:2008} of the displacement mechanism that incorporated multiphoton processes in the overlapping LL regime near $\eac=1$ does not extend to the fractional-$\eac$ regime.

In summary we have investigated the direct current response of fractional microwave induced resistance oscillations under variable microwave radiation power using a coplanar waveguide technique.
With the reasonable expectation that larger $N$ processes succeed small $N$ processes at larger microwave power, we find the simple, separated-Landau-level displacement description of Eq.\,\ref{eq} describes the observed $\edc$-frequency change.
Our results suggest that a theoretical treatment of combined dc excitation and microwave photons of arbitrary number when the levels are well separated is necessary.

We would like to thank M. A. Zudov for enlightening conversations and critical comments of the manuscript.
The work was supported  through DOE grant DE-FG02-05-ER46212 (NHMFL/FSU) and DE-SC0006671 (Purdue).   
The National High Magnetic Field Laboratory,  is supported by NSF Cooperative Agreement No. DMR-0654118, by the State of Florida, and by the DOE.

\bibliographystyle{apsrev}

\end{document}